# Emergence of localized plasticity and failure through shear banding during microcompression of a nanocrystalline alloy

**Amirhossein Khalajhedayati [a], Timothy J. Rupert [a,b,*]**

[a] Department of Chemical Engineering and Materials Science, University of California, Irvine, CA 92697, USA

[b] Department of Mechanical and Aerospace Engineering, University of California, Irvine, CA 92697, USA

[*] Corresponding author: Tel: +1 949 824 4937; Email Address: trupert@uci.edu (T.J. Rupert).

**Abstract**

Microcompression testing is used to probe the uniaxial stress-strain response of a nanocrystalline alloy, with an emphasis on exploring how grain size and grain boundary relaxation state impact the complete flow curve and failure behavior. The yield strength, strain hardening, strain-to-failure, and failure mode of nanocrystalline Ni-W films with mean grain sizes of 5, 15, and 90 nm are studied using taper-free micropillars that are large enough to avoid extrinsic size effects. Strengthening is observed with grain refinement, but catastrophic failure through strain localization is found as well. Shear banding is found to cause failure, resembling the deformation of metallic glasses. Finally, we study the influence of grain boundary state by employing heat treatments that relax nonequilibrium boundary structure but leave grain size unchanged. A pronounced strengthening effect and increased strain localization is observed after relaxation in the finer grained samples.

## 1. Introduction

Nanocrystalline metals, commonly defined as polycrystals with a mean grain size (*d*) of less than 100 nm, are promising structural materials [1], mainly due to reports of high strength [2,3], fatigue resistance [4,5], and wear resistance [6–8]. When grain size is reduced below ~100 nm, new physical mechanisms begin to carry plastic deformation. First, there is a shift to plasticity that is controlled by grain boundary sites acting as sources and sinks for dislocation activity. Van Swygenhoven et al. [9] used molecular dynamics simulations to study the deformation physics of nanocrystalline Al and found that dislocation nucleation and propagation was limited by activity at grain boundary sites. Interestingly, they found that these dislocations often became pinned at grain boundary ledges as they moved through the grain, giving the spacing between boundary pinning points as the characteristic length scale of the mechanism. This interpretation has been supported by Huang et al. [10], who showed that experimental data could be well-described by a model that invokes Orowan-type pinning of dislocations with the grain size taken as the distance between obstacles. The behavior of nanocrystalline materials with grain sizes blow ~10-20 nm has been attributed to the emergence of grain boundary sliding and rotation as the dominant carriers of plastic deformation. Schiotz et al. [11,12] were the first to report such a mechanism when they detected local sliding events during molecular dynamics simulations of nanocrystalline Cu.

The common feature of the new deformation physics described above is the increased importance of grain boundaries as facilitators for plastic deformation. Since grain boundaries are more abundant and more important in nanocrystalline systems, increased attention has been focused on studying the atomic structure of these interfaces. A number of studies have reported that nanocrystalline metals often contain nonequilibrium grain boundaries, characterized by excess free volume or grain boundary dislocations, in their as-prepared state [13,14]. However, this nonequilibrium structure can be easily relaxed by applying low temperature heat treatments [13,15], with a more ordered and connected grain boundary structure found after relaxation. Perhaps not surprisingly since boundaries are so important for nanocrystalline plasticity, reports have shown that mechanical strength is highly dependent on this grain boundary structural state. Detor and Schuh [16] showed that grain boundary relaxation resulted in a significant increase in hardness, even though grain size was unchanged. Rupert et al. [17] further isolated this effect through systematic nanoindentation at different grain sizes, showing that this hardening occurred quickly and was grain size dependent.

The discussion above highlights the fact that novel deformation physics control plasticity in nanocrystalline materials and shows that these mechanisms are sensitive to grain boundary state. However, the vast majority of studies which probe mechanical behavior systematically as a function of grain size or grain boundary state rely on indentation experiments (e.g., [16–21]). While such techniques allow for large numbers of tests and only require small volumes of material, they also only give a scalar measurement of strength. As a result, the community has very little information about how the novel deformation mechanisms in nanocrystalline materials impact more complex behavior, like strain hardening and failure. In addition, nanoindentation imposes a complex, three-dimensional stress state which makes it difficult to connect with constitutive theories for yielding and also adds concerns about the effect of a large confining pressure on plasticity. In order to study strain hardening behavior, full plastic flow, and the failure of nanocrystalline materials in a straightforward manner, a simple uniaxial tension or compression test is required. However, such testing has to date been problematic as premature failure can occur for two main reasons: (1) improper specimen geometry, and (2) processing defects.

Early attempts at uniaxial testing of nanocrystalline materials largely consisted of creating dog-bone specimens from thin sheets of nanocrystalline metals (e.g., [22–24]). This means that the samples commonly had thicknesses that were orders of magnitude smaller than the in-plane dimensions. Such geometries are problematic, as they introduce a geometric sample size effect, with strain-to-failure decreasing as sample thickness decreases [25]. Brooks et al. [26] explored this effect specifically in nanocrystalline Ni, showing that samples with thicknesses below ~100 µm experienced macroscopically brittle fracture that was not representative of the intrinsic material response. Zhao et al. [27] showed that by making sample geometries defined by ASTM Standard E8 [28], the strain-to-failure becomes independent of thickness, leading to a recommendation that the comparison of strain-to-failure measurements of a nanocrystalline material taken from different tensile test specimen geometries should be done cautiously.

Another complication that precludes the simple application of traditional uniaxial testing techniques to nanocrystalline materials is the effect of processing defects. Research has shown that nanocrystalline materials are commonly plagued by incomplete consolidation of particles, surface flaws, sulfur-induced grain boundary embrittlement, and hydrogen pitting, all of which can cause premature failure [23,29,30]. For example, nanocrystalline Cu [31] and Ni-Fe [30] showed increased strain-to-failure with improved processing chemistry that reduced particulate contamination and hydrogen pitting. Brooks et al. [26] also showed that nanocrystalline Ni samples produced by an optimized process experienced twice as much plastic strain before failure, while the samples without this optimization always failed at large void-like defects produced when hydrogen gas was trapped in the deposit. Therefore, without having a proper geometry for mechanical testing and samples that are free of processing defects, conventional testing methods cannot provide us with accurate results and an alternative uniaxial testing technique is needed to adequately probe the plastic flow and failure response of nanocrystalline materials.

Recently, microcompression testing has become a promising and reliable technique that can be used to acquire the mechanical properties from a small volume of material [32,33]. Although this testing method is often used to study the effects of external sample size on mechanical behavior (e.g., [34]), such micropillars can actually serve as a bulk mechanical testing technique if the characteristic length scale associated with the microstructure of the material is much smaller than the pillar size. For a material with a grain size in the nanometer range, hundreds of thousands to millions of crystallites will be contained inside of a pillar with a diameter of at least a few microns. For this reason, we suggest that micropillar compression can be used to measure the intrinsic properties of nanocrystalline materials. With microcompression, one can use sample aspect ratios that are small and within the range of ASTM standards while also minimizing the possibility of processing voids and defects being trapped in the small volume of material that is probed.

In this paper, we use uniaxial microcompression testing to study the full flow curve and failure behavior of a nanocrystalline alloy, with a specific focus on understanding the importance of grain size and grain boundary relaxation state. Nanocrystalline Ni-W was chosen as a model system, since grain size can be easily manipulated during electrodeposition and this system has been studied extensively with nanoindentation [16,35,36]. To the authors' best knowledge, this is the first study to systematically explore uniaxial flow and failure in specimens with grain sizes from near 100 nm to below 10 nm. By studying this entire range, we are able to probe the effects of the entire gamut of deformation physics that control plastic deformation in nanocrystalline metals, spanning grain boundary dislocation mechanisms as well as grain boundary sliding and rotation. Strength, strain hardening, strain-to-failure, and failure mode are tabulated as a function of grain size and relaxation state. We show that reducing grain size and relaxing grain boundary structure can change the failure mode of a nanocrystalline metal from uniform plastic flow to shear localization. The localized plastic flow that we observe is reminiscent of shear banding in metallic glass, providing clear evidence of a connection between the deformation physics of nanocrystalline and amorphous metals.

## 2. Materials and Methods

Nanocrystalline Ni-W alloy samples were created using pulsed electrodeposition following the work of Detor and Schuh [36]. In this system, the applied waveform is used to control the W content in the deposited film and, since the W exhibits a subtle tendency to segregate to the grain boundaries, the grain size can be controlled as well. Samples with mean grain sizes of 5, 15, and 90 nm were deposited onto 99% pure Ni substrates. In order to limit the impurities in our films, no organic grain refiner such as saccharine was used during electrodeposition. After deposition, samples were divided into two sets: as-deposited and relaxed. To relax the nonequilibrium grain boundary structure found in electrodeposited films, the specimens were annealed at 300°C for 1 hour and then water quenched, following prior work by Rupert et al. [17]. Such low temperature annealing treatments are aggressive enough to relax the grain boundaries, but gentle enough that grain growth is not induced. In light of recent computational [37] and experimental [38] research connecting grain boundary chemistry to strength, one could also ask if segregation of additional solutes to the grain boundaries occurs during annealing. Fortunately, previous microstructural evolution studies of nanocrystalline Ni-W [16] have shown that grain boundary relaxation always precedes other types of structural evolution (precipitation of second phases, short-range ordering, or segregation), making Ni-W an ideal candidate for isolating the effects of nonequilibrium grain boundary structure. All of the samples were polished to submicron level using diamond suspension solutions. The final thicknesses of the films were at least 50 µm after fine polishing.

It is important to reiterate that grain size and composition are not independent in the Ni-W system. Films with higher W content will have finer grain sizes, which can complicate direct comparisons of strength measurements from different grain sizes. Rupert et al. [39] were able to isolate and quantify the strengthening effect of solid solution W addition to nanocrystalline Ni when grain boundary dislocations control plasticity, but the authors are not aware of any study which has isolated solid solution strengthening when grain boundary sliding dominates. This means that any trends we observed in measured strength cannot be attributed solely to grain size, nor can the relative contribution of grain size and composition be separated. However, Rupert et al. [39] also found that solid solution addition does not alter the dominant deformation physics. Since we are mainly interested in reporting on flow and failure characteristics in this paper, the three selected grain sizes can still be used to demonstrate mechanical behavior when the different deformation mechanisms are active.

Transmission electron microscopy (TEM) samples were prepared using the Focused Ion Beam (FIB) in-situ lift out technique [40] in a Quanta 3D FEG Dual Beam microscope. Lamellae with 10 µm widths and 5 µm heights were cut from each sample, and then thinned to create an electron-transparent specimen. A voltage of 5 kV was used during the final thinning step to minimize the thickness of the damaged layer created by the FIB. The same instrument was used to take SEM images of the pillar before and after deformation at 5-10 kV. Bright field TEM images were taken using a FEI/Philips CM-20 TEM microscope operating at 200 kV. Mean grain sizes were calculated by manually identifying and tracing individual grains, and then calculating the equivalent circular diameter for each before finding an average value. Fig. 1(a)-(c) shows the bright field TEM images of the as-deposited samples, while Fig. 1(d) presents cumulative distribution functions of grain size in each sample. Fig. 1 shows equiaxed grains with a narrow grain size distribution. X-ray diffraction (XRD) profiles were obtained using a Rigaku SmartLab X-ray Diffractometer with a Cu Kα radiation source operated at 40 kV and 44 mA. The XRD profiles were used to ensure that all of the specimens were polycrystalline, fcc solid solutions and to estimate the average grain size using the Scherrer equation [41].

Micropillars with average diameters of 7 ± 0.1 µm and average heights of 15.7 ± 0.3 µm were fabricated with automated lathe milling using the FIB, following the method of Uchic et al. [32]. The pillar aspect ratio (height/diameter) of 2.2 was chosen to follow microcompression testing guidelines developed by Zhang et al. [42], in order to avoid plastic buckling. The number of grains across the diameter of an average pillar is ~80 for the largest grain size ($d$ = 90 nm) sample, which is more than enough to avoid external size effects on strength [43]. To make sure the indenter head does not hit the surrounding material around the pillar, a crater with a diameter of ~60 µm was first milled using 7-15 nA around the prospective site. A circular fiducial mark was then milled on the center of the rough pillar with a 100 pA current, and a two-step lathe milling process was utilized to create the final pillar shape. First, a 1 nA current was used to mill the pillar close to its final shape, and then a final current of 100 pA was used to create a higher quality surface on the pillar and to reduce the damage layer from the FIB [44,45]. The lathe milling method allows taper-free pillars to be produced, as shown in Fig. 2, so a simple, uniform stress state can be applied to the nanocrystalline specimens.

An Agilent G200 nanoindenter was used to perform microcompression testing. The pillars were compressed using a Berkovich indenter tip truncated to have a flat, triangular surface at the end with an average side length of 31 µm. Although the instrument is an inherently force-controlled indenter, a feedback control loop was implemented so that the indenter nominally imposes a constant displacement rate. A constant displacement rate of 5 nm/s was applied, resulting in an engineering strain rate of $3.2 \times 10^{-4}$ s$^{-1}$ for the samples used here. For each grain size and grain boundary relaxation state, at least five pillars were tested to ensure that the results were reproducible. When using microcompression testing, one must be careful to correct for the compliance of the base and minimize the misalignment between the indenter head and the pillar's top surface [42,46,47]. Sneddon's equation of rigid cylindrical flat punch displacement into an elastic half space was used to exclude the base compliance from our data [46]. Misalignment of the indenter decreases the measured modulus of the pillar by introducing a bending component to the deformation [42]. By carefully preparing flat, level samples and by following the best practices to align the pillars with the nanoindenter head [32], we are able to limit the misalignment to less than 1°, as calculated using the method of Schuster et al. [48].

Engineering stress-strain data was converted to true stress-stain data by assuming a uniform plastic deformation and conservation of the volume of pillar. Brandstetter et al. [49] and Vo et al. [50] both showed that the traditional 0.2% offset yield strength is not appropriate for nanocrystalline materials. Therefore, following the work of Brandstetter et al., yield stress was calculated based on a 0.7% yield strain offset. Some pillars were unloaded in order to calculate the modulus of the specimens while others were stopped at different plastic strains to study the deformation behavior in the SEM. To compare the results of microcompression with standard nanoindentation techniques, a Berkovich tip calibrated with standard fused silica was used for the measurement of hardness for each specimen. A constant indentation strain rate of 0.05 s$^{-1}$ and at least 30 indentation experiments were used to find average hardness values.

## 3. Effect of grain size: As-deposited nanocrystalline Ni-W

We begin by exploring how grain size and, therefore, the dominant physical mechanism controlling plasticity affects the uniaxial deformation of nanocrystalline materials. The $d = 5$ nm sample is small enough that grain boundary sliding and rotation is expected to completely control plastic deformation. The $d = 90$ nm is near the upper limit of the nanocrystalline regime and the deformation physics is expected to be dominated by dislocation activity originating at and being absorbed by the grain boundaries. The $d = 15$ nm is in the critical grain size range mentioned in the introduction where the shift between these two mechanisms occurs, so some combination of grain boundary dislocation plasticity and grain boundary sliding/rotation is expected here. Fig. 3 presents true stress-strain curves for as-deposited Ni-W specimens with different grain sizes.

The results for the $d = 90$ nm pillars, presented in Fig. 3(a), are discussed first. There is little variation between individual sample response and an average yield stress of 1.54 GPa is measured. Immediately after yielding, very limited strain hardening is observed over a small range of plastic strain. However, the vast majority of the plastic behavior is characterized by subtle strain softening. The decrease in flow stress with plastic strain is roughly linear, with a measured slope -0.45 GPa. The 90 nm grain size material lies near the upper limit of what is considered nanocrystalline ($d < 100$ nm), and a similar strain softening behavior has been previously observed in ultra-fine grained Ti [51] and Ni [52]. A possible reason for the observed decrease in strain hardening ability compared to traditional metals is the reduction of dislocation sources inside the grain, and hence the reduction of hardening mechanisms expected from these activities. Fig. 3(a) also shows that the sample with $d = 90$ nm can sustain large plastic strains without any signs of failure, even after an applied strain of >25%. Fig. 4(a) shows an SEM image of a $d = 90$ nm pillar after 30% compressive strain. Plastic deformation was found to be uniform throughout the sample and no significant bending of the pillar was observed.

Fig. 3(b) shows the true stress-strain data for the 15 nm grain size material, and several aspects of the stress-strain behavior have changed significantly with grain refinement. First, the average yield stress has significantly increased to 2.57 GPa. Such an increase in strength with grain refinement mimics reports that rely on nanoindentation results [17]. The strain hardening behavior during the initial stages of plastic deformation appears to be similar for the 90 nm and 15 nm grain sizes; a small amount of strain hardening is observed immediately after yielding but this behavior is short-lived. However, the strain softening at larger plastic strains is much more pronounced and a linear strain softening slope of -1.95 GPa is measured for this grain size. Perhaps the most dramatic and unexpected difference in the mechanical behavior of the $d = 15$ nm samples is the observation of sudden failure at true strains of ~10-20%. Fig. 4(b) shows an SEM image of a $d = 15$ nm pillar after failure under compression. The pillar fails through the formation of a major shear band, and the top of the pillar shears off at an angle. Since the nanoindenter used in this study is inherently load controlled, this sudden downward movement of the indenter cannot be accommodated quickly enough in the feedback loop, causing a violent downward motion of the indenter tip that then pushes the damaged base off to the side. The grain size of this sample is close to the critical grain size where grain boundary sliding and rotation begin to carry an appreciable fraction of plastic strain [53,54], suggesting a causal link between the emergence of such collective grain boundary mechanisms and strain localization.

Finally, Fig. 3(c) presents the true stress-strain curves from the $d = 5$ nm specimens, showing even more unique mechanical behavior. These specimens demonstrate what appears to be elastic-perfectly plastic response with no discernible strain hardening and they then fail catastrophically after small plastic strains of only a few percent. The yield strength increases to 3.0 GPa, but the strain-to-failure has decreased significantly. No strain hardening or strain softening observed. While there appears to be increased scatter in the loading behavior of these samples, it is important to point out that four curves almost exactly overlap and are difficult to differentiate. The curves from the two samples that appear to be more compliant during loading (green and dark red in Fig. 3(c)) actually experience strain jumps that cause an apparent deviation from linear elasticity. These serrations are shown more clearly in the inset to Fig. 3(c). The formation and development of shear banding can be seen in SEM images taken after uniaxial compression. Fig. 5(a) shows a $d = 5$ nm pillar where the test was stopped immediately after the initial deviation from linear elastic loading in the load-displacement curve, showing that several shear bands have formed along the length of the pillar. Fig. 5(b) is a pillar from the same grain size sample, but it was allowed to fail completely. Catastrophic failure occurred and secondary shear bands can be seen crossing the main slip steps. The details of the surface relief resulting from the shear banding are seen more clearly in Fig. 5(c), which presents an SEM image taken at higher magnification. We hypothesize that the premature flow serrations shown in the inset to Fig. 3(c) are shear bands which could not fully cross the sample width and lead to complete failure. As mentioned previously, it is expected that grain boundary sliding and rotation are the main carriers of plastic deformation at $d = 5$ nm.

Further insight into the failure behavior of our $d = 5$ and $d = 15$ nm samples can be found by looking more closely at the indenter displacement. Fig. 6(a) presents the average displacement rate versus time for representative examples taken from each grain size. Two major types of displacement bursts can be identified. For the two finest grain sizes, the average displacement rate shows a discontinuity corresponding to shear banding and catastrophic failure of the pillar. Since the system is inherently force controlled, the lack of resistance due to sample failure leads to a rapidly increasing displacement rate. For the largest grain size, small strain bursts are observed, but the average displacement rate settles back to the constant target value. Here, dramatic changes in the slope of the stress-strain response give rise to the temporary strain bursts. For this sample, the change in the instantaneous sample stiffness is simply the result of a shift from elastic loading with a high slope to plastic flow with a much lower tangent modulus. After the yield point, while the indenter tries to keep the displacement rate constant, the feedback loop overcompensates and several strain bursts appear. The responses described above can also be seen in the raw displacement versus time data presented in Fig. 6(b). Shear banding leads to a rapid, uncontrollable failure of the pillar and the indenter then slams into the surface, while the behavior of the strain bursts in the $d = 90$ nm sample are much more controlled.

The strain localization described above for our 5 nm grain size is qualitatively similar to observations of shear banding by Trelewicz and Schuh [19] during nanoindentation studies of nanocrystalline Ni-W with $d = 3$ nm, although here we see more clearly how such shear banding reduces strain-to-failure. Trelewicz and Schuh observed the formation of shear bands in the pile-up surrounding indentation impressions created with a cube corner indenter tip, suggesting that the finest nanocrystalline metals deform through mechanisms that are similar to those that control metallic glass plasticity. The fundamental unit of plasticity in a metallic glass is a shear transformation zone (STZ), or a small collection of atoms that undergoes an inelastic shear strain. When metallic glasses are deformed at temperatures well below their glass transition temperature, STZ activity leads to free volume accumulation along a distinct path, causing strain localization in the form of shear banding [55]. Schuster et al. [48,56] studied room temperature plastic deformation of a Pd-based metallic glass using a microcompression technique similar to the one used here. These authors observed features of metallic glass plasticity that are similar to the behavior of our $d$ = 5 nm Ni-W sample: (1) serrated flow characterized by displacement bursts, (2) a lack of any appreciable strain hardening, and (3) failure through shear banding.

Insight into the similarity between the shear banding observed in the finest nanocrystalline metals and metallic glasses can be gained by understanding exactly how grain boundary rotation and sliding are accommodated. Lund et al. [57] used molecular statics simulations to study nanocrystalline plasticity and found that these collective motions result from the local rearrangement of small groups of atoms within the grain boundaries, a process similar to the STZs found in metallic glasses. One important difference is that, while any group of atoms can participate in such collective rearrangement in a metallic glass since there is no long range order, only the grain boundary atoms can participate in such motion in a nanocrystalline system. Since the crystalline grain interiors cannot participate in STZ-like motion, localization should still be limited by the connectivity of the interfacial network. However, recent molecular dynamics simulations from Rupert [58] have shown that a grain boundary percolation path of high strain can be easily formed across a wire diameter if the grain size is small enough. Once this path is formed, strain intensifies in this region with progressive deformation due to the lack of hardening mechanisms.

While a natural comparison between amorphous plasticity and the collective grain boundary plasticity of the very finest nanocrystalline grain sizes can be made, a more complicated process is needed to describe why the 15 nm grain size would experience shear banding and strain localization since grain boundary dislocation plasticity is also important for this material. In fact, while Trelewicz and Schuh [19] saw shear banding during quasi-static nanoindentation of a 3 nm grain size, they did not observe such behavior for larger grain sizes. To the best of our knowledge, the results presented here are the first to show shear banding in a nanocrystalline fcc metal with a grain size as large as 15 nm. With a grain size this large, a shear band is unlikely to form entirely through a grain boundary path, but rather through a combination of boundary and dislocation mechanisms. Hasnaoui et al. [59] provided the first evidence of such a mechanism, when their molecular dynamics simulations of nanocrystalline Ni showed the formation of common shear planes during uniaxial deformation at elevated temperatures. They suggested that this shear plane formation was a cooperative process that included several grains and could involve three types of mechanisms: migration of grain boundaries, coalescence of grains with low angle grain boundaries, and intragranular slip that provides continuation of a shear plane encountering a triple junction. Sansoz and Dupont [60] also observed the initial formation of a shear plane during molecular statics simulations of nanoindentation in nanocrystalline Al. Finally, Rupert [58] used molecular dynamics simulations of nanocrystalline Ni to show more clearly how a localization path formed by grain boundary and dislocation mechanisms can thicken and intensify to form a fully developed shear band. This author observed successive partial dislocation emission, leading to deformation twins which extended the localization path through the grain interior. Rupert also observed that, while localization paths along grain boundaries form quickly during initial stages of plasticity, grains that must be sheared with grain boundary dislocations can limit shear band formation. This likely explains why the 15 nm grain size sample experiences a moderate level of plastic strain before shear banding. Until dislocation mechanisms can traverse enough grains to create a percolation path across the sample width, catastrophic strain localization cannot commence.

**4. Effect of grain boundary state: Relaxed nanocrystalline Ni-W**

With insight into the importance of grain size, we next move our attention to understanding how grain boundary state, specifically the relaxation of nonequilibrium interfacial structure, affects the mechanical properties of nanocrystalline alloys. XRD and TEM of the annealed samples showed that the low temperature annealing treatment used for relaxation did not change the grain size. We began first by looking at the largest grain size sample. Fig. 7(a) shows the true compressive stress-strain curves for the $d = 90$ nm samples. The yield strength of the material has slightly increased to 1.7 GPa, or 0.15 GPa more than the as-deposited case. We again observe two different types of behavior in the true stress-strain curve after the yield point. First, a very limited amount of strain hardening is found immediately after yielding over a range of small plastic strains. This is followed by a region of roughly linear softening behavior, but this behavior has become slightly more pronounced and was measured to have a slope of -0.63 GPa. Hence, even large nanocrystalline grain sizes become more prone to strain softening after grain boundary relaxation. Like its as-deposited counterpart, this material can still withstand large plastic strains of up to 25% without failing. SEM images of micropillars after deformation (not shown here) demonstrate that the relaxed, $d = 90$ nm grain size sample still experiences homogeneous plastic flow.

Fig. 7(b) shows the true stress-strain behavior of the $d = 15$ nm relaxed samples. The average yield stress of the material increased to 3.15 GPa, or increased by 0.57 GPa from the as-deposited state. Hence, grain boundary relaxation has a larger strengthening effect as grain size is reduced. Rupert et al. [17] suggested that relaxation reduces the density of grain boundary sources for dislocation emission or nucleation, making it harder to initiate plastic flow. Alternatively, Van Swygenhoven et al. [9] proposed that dislocation pinning at grain boundary sites could determine the strength of nanocrystalline metals, so it is possible that the local structure of the grain boundary influences this process as well. In this case, a more ordered (i.e., relaxed) grain boundary would have fewer high energy sites where local stress variations could aid the applied stress and keep the dislocation moving, again meaning a higher applied stress is needed to initiate macroscopic plasticity in the sample. The maximum flow stress and yield stress values do not differ more than the error in the measured data, indicating that even the limited strain hardening ability seen in the as-deposited sample has vanished. The value of the strain softening slope has decreased to -3.2 GPa for the relaxed $d = 15$ nm sample. Interestingly, the strain-to-failure of the relaxed sample decreased significantly compared to the as-deposited sample, with all of the pillars failing between 5% and 8% strain through shear banding. Thus, both $d = 90$ nm and $d = 15$ nm samples have a higher tendency to strain soften after relaxation, but the relaxed $d = 15$ nm sample also experiences catastrophic strain localization at smaller plastic strains when compared to the as-deposited state. Fig. 8(a) presents an SEM image of a $d = 15$ nm relaxed pillar right as it starts to fail, showing the formation of a major shear band.

Fig. 7(c) shows the true stress-strain behavior of the $d = 5$ nm relaxed specimens. For this group of specimens, no appreciable plastic strain was observed before catastrophic shear banding occurred. Without measurable plastic strain, we instead report the maximum stress as our yield strength here. The pillars fail as soon as they reach a critical stress of 3.4 GPa, meaning strength has increased by 0.4 GPa over the as-deposited condition. While this is appreciable strengthening, it is less than was observed for the 15 nm grain size, suggesting that relaxation has the strongest effect on strength at an intermediate grain size. Nanoindentation work from Rupert et al. [17] reported a trend that was qualitatively similar, but found that the largest strengthening was observed at $d = 6$ nm, near our finest grain size sample here. This apparent discrepancy can be addressed by directly comparing microcompression and nanoindentation measurements for the samples tested in this study, which we will do shortly. Fig. 8(b) shows a $d = 5$ nm relaxed pillar after the completion of the compression test, where it is clear that the pillar failed with a formation of a major shear band. The indenter head continued to travel downward for a few seconds after failure, causing the impression of the tip which is visible in the pillar top.

Fig. 9(a) shows our yield stress data plotted as a function of grain size for as-deposited and relaxed Ni-W, while Fig. 9(b) presents nanoindentation hardness measurements taken from the same samples. In each case, the strengthening we observed with grain refinement does not follow a strict $d^{1/2}$ scaling, signifying a deviation from Hall-Petch behavior. While compositional changes should also affect this data, the increase in W content with decreasing grain size should lead to strengthening, so it cannot be to blame for the data lying below a Hall-Petch trend in Fig. 9. Both sets of measurements show that relaxation of nonequilibrium grain boundaries leads to significant strengthening, but the maximum effect occurs at different grain sizes for each type of measurement. While yield stress demonstrates the largest improvement at $d = 15$ nm, hardness increases the most with relaxation for the 5 nm grain size. However, we suggest that this is simply an artifact associated with the nanoindentation technique. The stress-strain curves from the relaxed $d = 5$ nm sample show that this material cannot accommodate appreciable plastic strain before failing catastrophically through shear banding. However, during nanoindentation, either the confining pressure underneath the indenter or the geometry of the indenter tip can suppress such localization, leading to an anomalously large measurement of the sample's strength. This concept can be seen more clearly in Fig. 10, where nanoindentation hardness is plotted versus yield stress measurements. For all of our as-deposited samples and the relaxed samples with larger grain sizes (i.e., the specimens that can sustain at least a few percent plastic strain before failure), there is a linear correlation between hardness and yield stress. This is the conceptual idea behind hardness as a quick and convenient measure of strength: the two quantities should be related in a known manner. However, the relaxed $d = 5$ nm sample (i.e., where there is no measurable plastic strain before failure) clearly deviates from this behavior, with a strength that is much lower than what would be predicted from its hardness. For materials where strain localization occurs before stable plastic flow can be developed, such as our smallest nanocrystalline grain size in the relaxed state, hardness trends do not necessarily mimic trends in true strength measurements.

An important observation is that relaxation of nanocrystalline grain boundaries makes the 5 and 15 nm grain sizes more susceptible to shear banding, reducing the strain-to-failure in both cases. For $d = 5$ nm, a comparison with metallic glass physics can again be useful. Both experiments [61] and simulations [62] have shown that metallic glasses with increased short range order deform through larger, more conspicuous shear bands. A highly disordered metallic glass has many sites with elevated local stresses, which can cause small shear bands to nucleate in different regions of the sample and give a macroscopic deformation that is more homogeneous. When there are few local variations in atomic stress due to structural disorder, the operation of an STZ provides the largest local stress fluctuation and strongly biases successive nearby STZ operation, leading to large catastrophic shear bands. If deformation is accommodated entirely through grain boundary processes in the 5 nm grain size sample and these boundaries locally deform in a manner that is similar to a glass, a relaxed, more ordered grain boundary structure would also result in increased strain localization. For the 15 nm grain size, increased strain localization likely results from a superposition of the effect of relaxation on grain boundary processes and its effect on grain boundary dislocation mechanisms. As mentioned in the Introduction, grain boundaries not only act as sources of dislocations, but also as sinks where the defects are reabsorbed into the opposite boundary. A more ordered grain boundary, with fewer local stress variations and less free volume, should be a less efficient sink for such absorption than the disordered boundary found in the as-deposited materials. Without efficient reabsorption of the first dislocation that traverses the grain, a bias for successive nearby dislocation emission exists within the crystallite, making it easier for the localization path to be created across the grain interior.

The results described here highlight the fact that grain boundary state is very important for nanocrystalline metals. The grain size of a nanocrystalline metal is often thought of as the structural feature which controls mechanical properties, but we show here that the plastic flow and failure of nanocrystalline Ni-W samples with the same grain sizes can be dramatically different depending on their grain boundary relaxation state. A more ordered boundary structure gives nanocrystalline metals increased strength, but also leads to more pronounced strain softening during the later stages of plastic deformation. In addition, for our smallest grain size of 5 nm and 15 nm, relaxation of nonequilibrium grain boundaries reduces strain-to-failure and promotes shear banding. While an ordered boundary structure improves strength, this appears to come at the expense of toughness.

Finally, since grain boundary state has been shown to be important, it is likely necessary to differentiate between nanocrystalline materials created by different processing routes when looking for trends in literature data. Early studies of nanocrystalline materials often accessed a variety of grain sizes by taking a very fine grained sample and annealing it to cause thermal grain growth (see, e.g., [63,64]). However, such annealing should also relax nonequilibrium grain boundary structure, making it difficult to compare property measurements from, for example, a 40 nm grain size sample that was created by annealing with those from an as-deposited 40 nm sample. Grain boundary state should also be important when comparing deposited nanocrystalline materials with those created by severe plastic deformation. Processing techniques such as ball milling, high pressure torsion (HPT), and equal channel angular pressing (ECAP) create nanostructured metals by adding a great deal of strain energy to the material in order to drive refinement. Materials created by these methods likely have grain boundary structures which are even further from equilibrium than deposited films.

## 5. Conclusions

Microcompression testing has been used to study the effects of grain size and grain boundary relaxation state on plastic flow and failure of nanocrystalline Ni-W. To our knowledge, this is the first study to systematically probe the uniaxial stress-strain response of nanocrystalline metals with microcompression over a range of grain sizes that spans the entire range of possible deformation mechanisms. Such a technique is extremely advantageous for probing nanostructured materials, as it avoids the common geometric and processing artifacts which plague standard uniaxial testing on these materials. The results presented here allow the following conclusions to be drawn:

- Grain refinement from $d = 90$ nm to $d = 5$ nm causes yield strength to nearly double, increasing from 1.54 GPa to 3.0 GPa, respectively, although chemistry also plays a role in this strengthening since W content increases from ~3% to ~20%.

- Our largest grain size, $d = 90$ nm, could be compressed to >25% true strain without failing. Subtle strain softening was observed, but deformation remained homogenous in nature throughout the compression experiments.

- The intermediate grain size of 15 nm, where a combination of dislocation plasticity and grain boundary plasticity controls deformation, was much stronger as a result of its finer grain structure, but experienced pronounced strain softening and then sudden failure through shear banding at applied true strains of 10-20%.

- Our finest grain size, $d = 5$ nm, exhibits elastic-perfectly plastic deformation with no apparent strain hardening or softening after yield. After plastic strains of only a few percent, strain localization occurs and these specimens fail through shear banding that resembles the behavior of metallic glasses.

- Relaxation of nonequilibrium grain structure strengthens nanocrystalline metals, but makes our two finest grain sizes more susceptible to strain localization. For $d = 15$ nm, the strain-to-failure is reduced to 5-8%, while the 5 nm grain size sample shows no appreciable plastic strain before shear banding causes failure.

- In samples where strain localization leads to failure, there may not be a direct correlation between nanoindentation hardness and yield stress measurements. For our relaxed $d = 5$ nm samples, hardness was artificially high due to the suppression of shear banding under the indenter.

Taken as a whole, the results presented here show that catastrophic strain localization is an issue for nanocrystalline metals with small grain sizes. The grain sizes less than 20 nm which are strongest are also the most likely to fail at small applied strains. This strain localization is also a function of grain boundary state, with an ordered interfacial structure promoting shear banding and failure that resembles metallic glass behavior. While a relaxed grain boundary state increases strength, it is detrimental to strain-to-failure.

**Acknowledgments**

This research was supported primarily by the US Army Research Office through Grant W911NF-12-1-0511. T.J.R. acknowledges partial additional support from the Broadening Participation Research Initiation Grants in Engineering (BRIGE) program from the National Science Foundation under Grant CMMI-1227759.

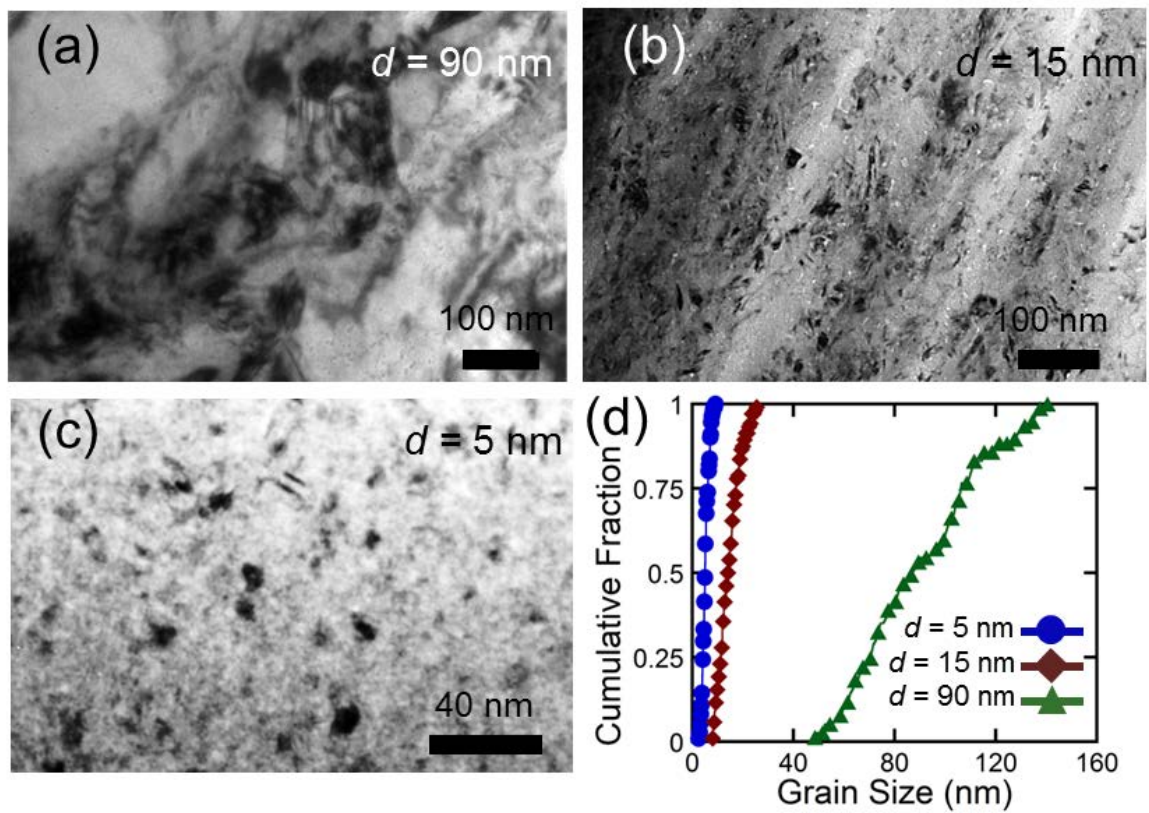

**Fig. 1.** Bright field TEM images of nanocrystalline Ni-W samples with average grain sizes of (a) 90 nm, (b) 15 nm, and (c) 5 nm. Grain size measurements from each sample are presented in a cumulative distribution plot in (d).

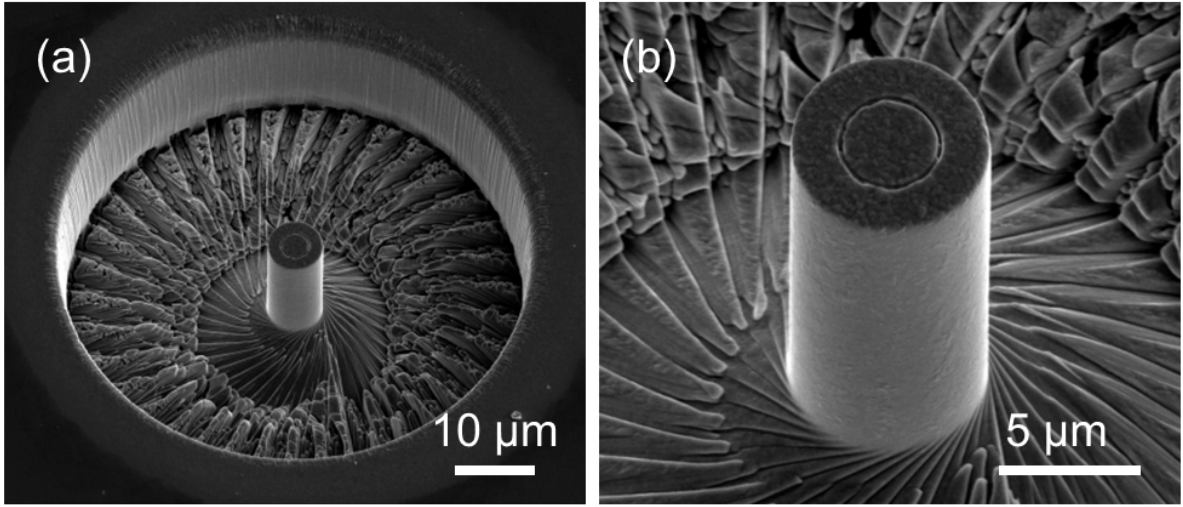

**Fig. 2.** (a) SEM image of a representative pillar that was used for uniaxial microcompression testing. (b) Magnified SEM image of the same pillar, showing the taper-free geometry.

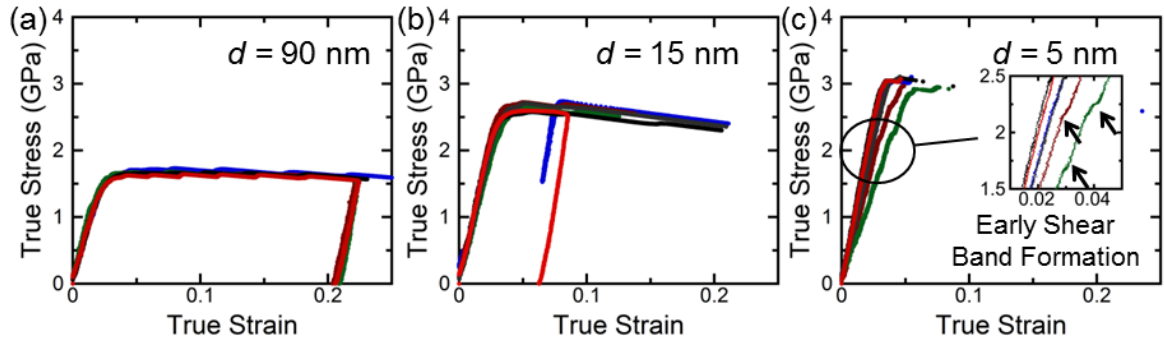

**Fig. 3.** True compressive stress-strain curves from as-deposited Ni-W samples with (a) $d$ = 90 nm, (b) $d$ = 15 nm, and (c) $d$ = 5 nm. At least 5 pillars were tested for each grain size, and some pillars were unloaded to calculate the stiffness of the pillar-substrate system. The inset to (c) shows that premature flow serrations can be found in a few of the curves from the 5 nm grain size samples.

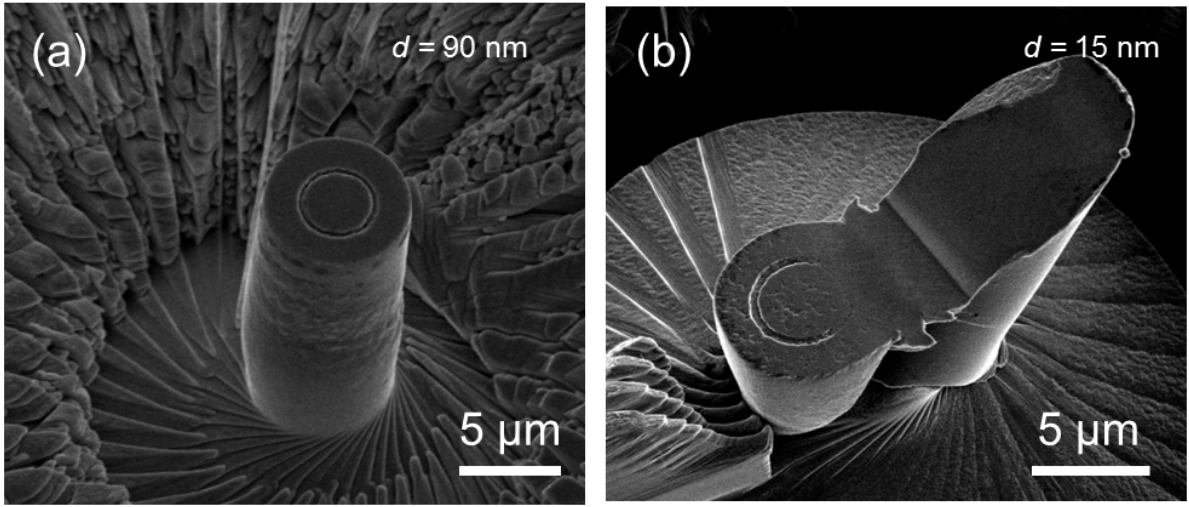

**Fig. 4.** SEM images of as-deposited nanocrystalline Ni-W alloy pillars with (a) *d* = 90 nm and (b) *d* = 15 nm after uniaxial microcompression. The 90 nm grain size pillar shows a uniform plastic deformation while the 15 nm grain size pillar fails through strain localization.

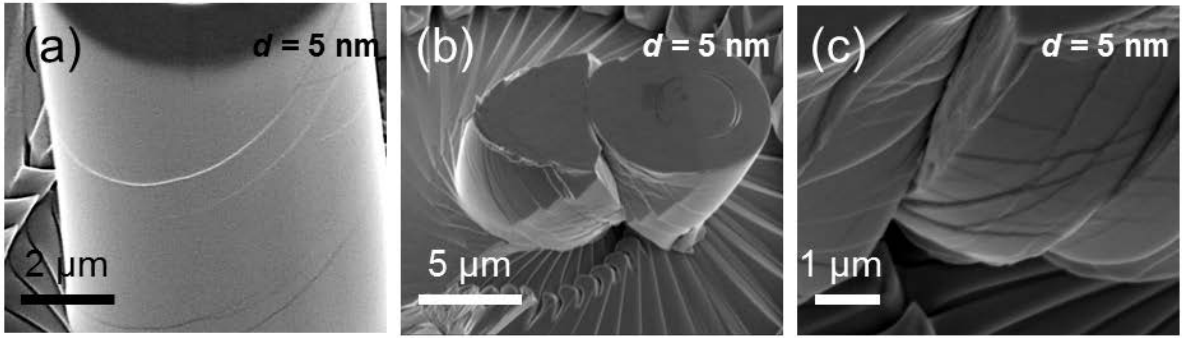

**Fig. 5.** (a) SEM image showing the formation of shear bands immediately after initial yield in as-deposited nanocrystalline Ni-W with *d* = 5 nm. (b) SEM image of the same material after complete failure showing catastrophic shear banding. (c) Magnified image of the same pillar showing intersecting shear bands in more detail.

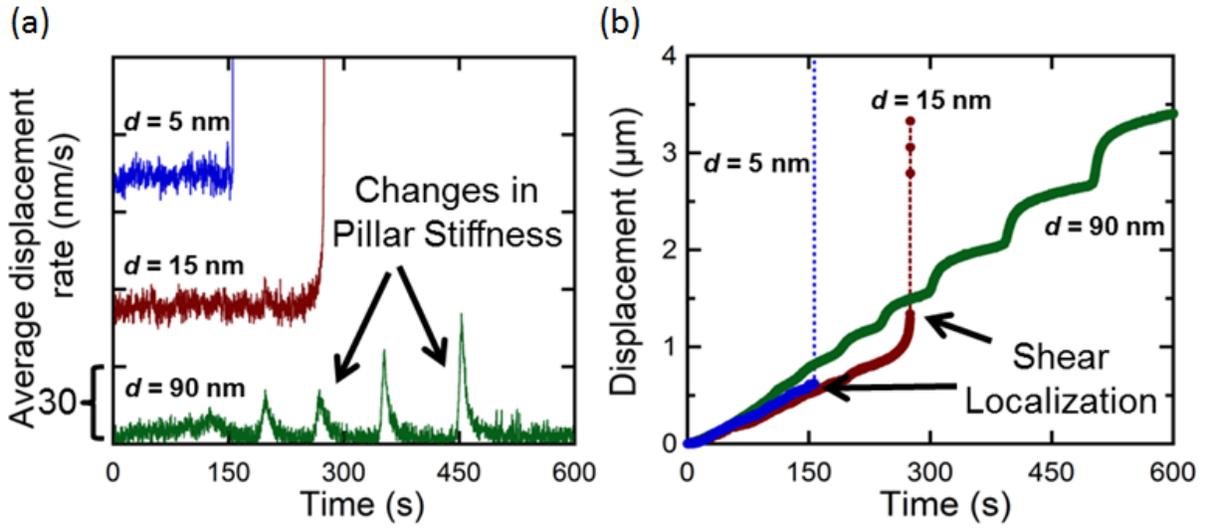

**Fig. 6.** (a) Average displacement rate of the indenter head during microcompression testing for a representative pillar from each grain size material. The average displacement rate target is 5 nm/s. Sudden strain burst are either due to shear localization or large changes to the stiffness of the pillar-substrate system. (b) The raw displacement versus time data also bears evidence of these strain bursts.

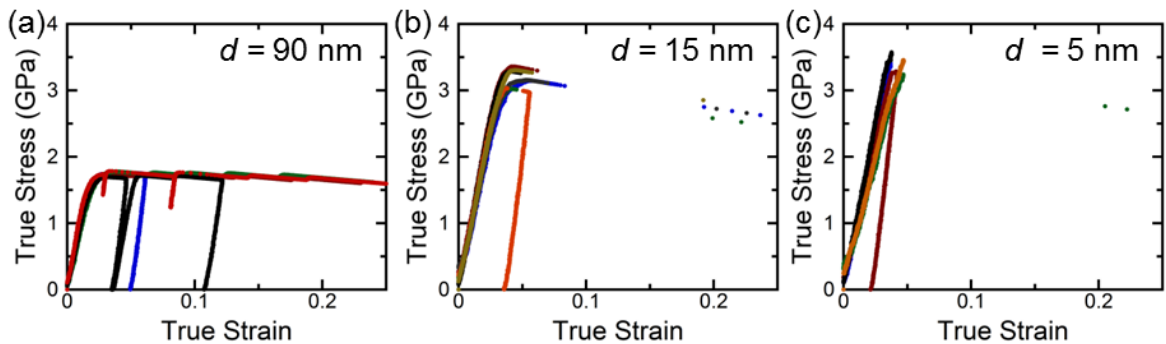

**Fig. 7.** True compressive stress-strain curves from relaxed Ni-W samples with (a) *d* = 90 nm, (b) *d* = 15 nm, and (c) *d* = 5 nm. At least 5 pillars were tested for each grain size, and some pillars were unloaded to calculate the stiffness of the pillar-substrate system.

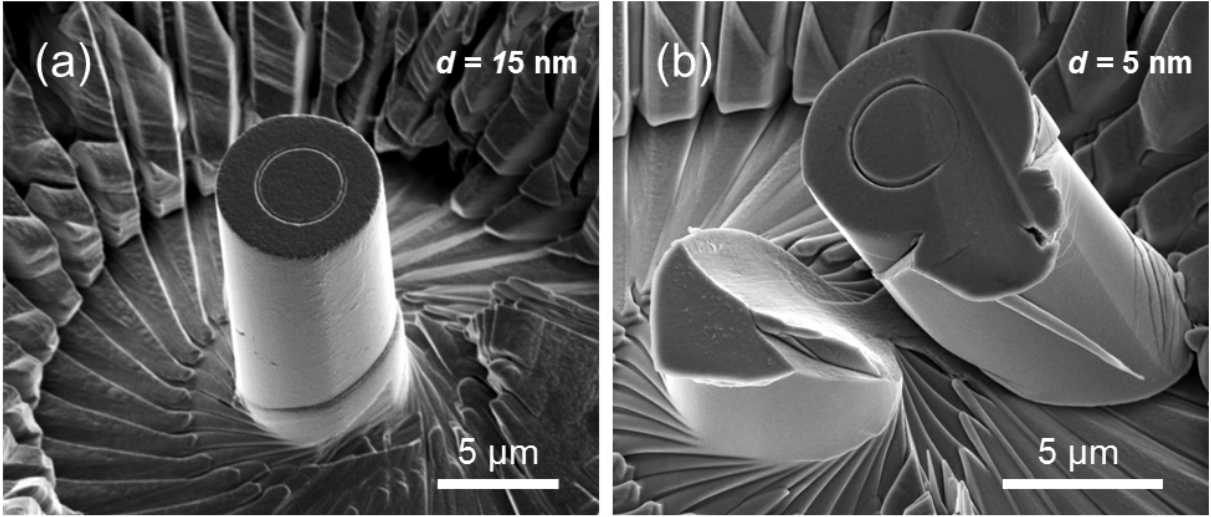

**Fig. 8.** SEM images of (a) $d$ = 15 nm and (b) $d$ = 5 nm relaxed samples after testing. After the complete failure of the pillar in (b), the indenter head pushes the pillar to the side and the impression of the indenter tip is visible.

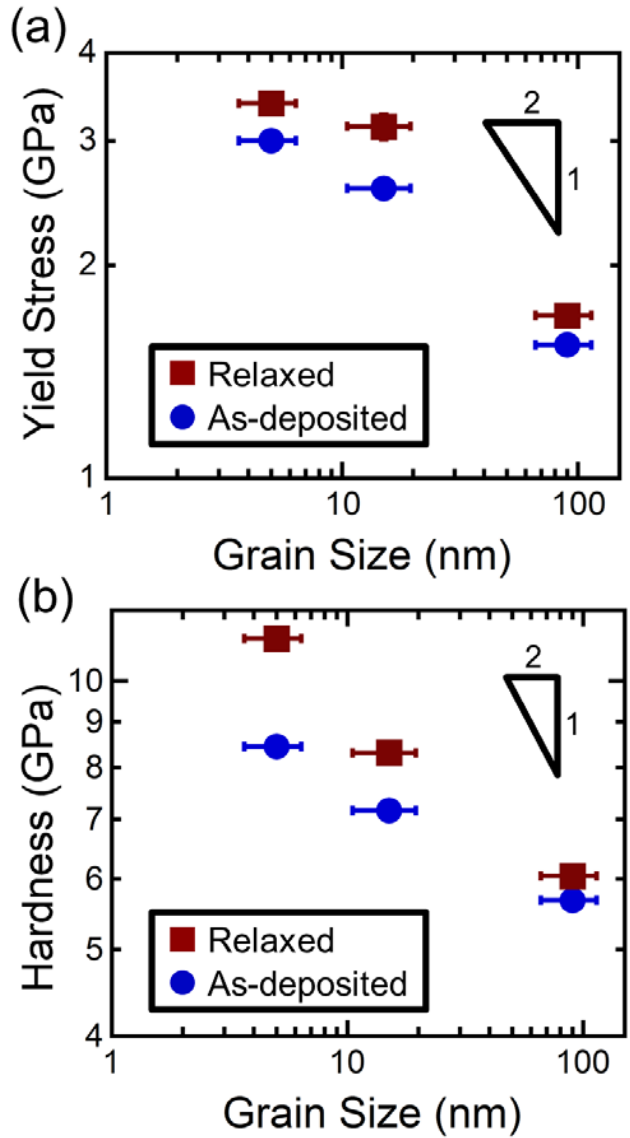

**Fig. 9.** Mechanical property measurements for nanocrystalline Ni-W samples, plotted as a function of grain size. Yield stress is shown in (a), while hardness measurements are presented in (b). For these samples, relaxation has a maximum effect on yield stress for $d$ = 15 nm, while hardness shows the largest increase for $d$ = 5 nm.

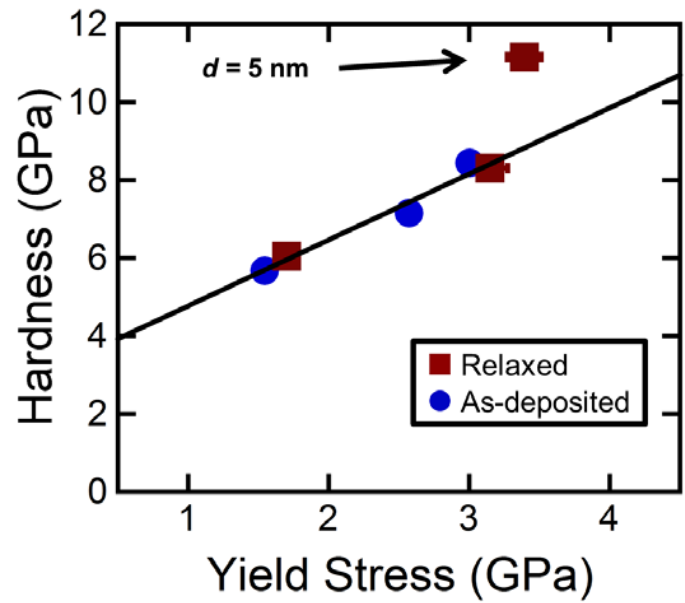

**Fig. 10.** Hardness plotted against yield stress for both as-deposited and relaxed nanocrystalline Ni-W samples. The straight line is fitted to the first five data points. The relaxed 5 nm grain size sample does not fall on this line, suggesting that hardness and yield stress cannot be directly related for nanocrystalline samples that fail before developing appreciable plastic strain.